\documentclass[sigconf,screen=true,anonymous=false,authorversion=true]{acmart}
\usepackage[utf8]{inputenc}
\usepackage{graphicx}
\graphicspath{{incl/}}     
\usepackage{setspace}

\copyrightyear{2019}
\acmYear{2019}
\setcopyright{acmcopyright}
\acmConference[DAC '19]{The 56th Annual Design Automation Conference 2019}{June 2--6, 2019}{Las Vegas, NV, USA}
\acmBooktitle{The 56th Annual Design Automation Conference 2019 (DAC '19), June 2--6, 2019, Las Vegas, NV, USA}
\acmPrice{15.00}
\acmDOI{10.1145/3316781.3317780}
\acmISBN{978-1-4503-6725-7/19/06}

\newtoggle{screen}
\toggletrue{screen}
\AtEndPreamble{%
\iftoggle{screen}
{
	\hypersetup{colorlinks, 
		linkcolor=blue,
		citecolor=blue,
		urlcolor=blue,
		filecolor=blue}
}
{
	\hypersetup{hidelinks}
}}

\usepackage[skip=8pt]{caption}            
\usepackage{cleveref}
\usepackage{multirow}
\usepackage{courier}
\usepackage{pgf}

\usepackage{pgfplots}
\usepackage{textgreek}
\usepackage[subrefformat=parens,labelformat=parens]{subfig}
\usepgfplotslibrary{groupplots}
\usetikzlibrary{patterns}

\usepackage{pifont}

\definecolor{CADENCE_VERY_DARK_GREY}{RGB}{60,60,60}
\definecolor{CADENCE_VERY_DARK_BLUE}{RGB}{9,105,141}
\definecolor{CADENCE_VERY_DARK_CYAN}{RGB}{9,158,156}
\definecolor{myorange}{RGB}{228,126,71}    
\definecolor{myblue}{RGB}{71,162,241}      
\definecolor{OMNI_MERCURY}{RGB}{225,225,225}        
\definecolor{OMNI_RICH_SKY_BLUE}{RGB}{71,162,241}   
\definecolor{OMNI_SALMON}{RGB}{247,120,138}         

\title[]{
Attacking Split Manufacturing\\
from a Deep Learning Perspective
}

\author{Haocheng Li}
\affiliation{
    \institution{CUHK}
}
\email{hcli@cse.cuhk.edu.hk}

\author{Satwik Patnaik}
\affiliation{
    \institution{NYU}
}
\email{sp4012@nyu.edu}

\author{Abhrajit Sengupta}
\affiliation{
    \institution{NYU}
}
\email{as9397@nyu.edu}

\author{Haoyu Yang}
\affiliation{
    \institution{CUHK}
}
\email{hyyang@cse.cuhk.edu.hk}

\author{Johann Knechtel}
\authornote{Four authors contributed equally to this research.}
\affiliation{
    \institution{NYU Abu Dhabi}
}
\email{johann@nyu.edu}

\author{Bei Yu}
\authornotemark[1]
\affiliation{
    \institution{CUHK}
}
\email{byu@cse.cuhk.edu.hk}

\author{Evangeline~F.~Y.~Young}
\authornotemark[1]
\affiliation{
    \institution{CUHK}
}
\email{fyyoung@cse.cuhk.edu.hk}

\author{Ozgur Sinanoglu}
\authornotemark[1]
\affiliation{
    \institution{NYU Abu Dhabi}
}
\email{ozgursin@nyu.edu}

\begin{document}

\begin{abstract}
The notion of integrated circuit split manufacturing
which delegates the front-end-of-line (FEOL) and back-end-of-line (BEOL) parts
to different foundries, is to prevent overproduction, piracy of the intellectual property (IP),
or targeted insertion of hardware Trojans by adversaries in the FEOL facility.
In this work, we challenge the security promise of split manufacturing
by formulating various layout-level placement and routing hints
as vector- and image-based features.
We construct a sophisticated deep neural network which can infer
the missing BEOL connections with high accuracy.
Compared with the publicly available network-flow attack~\cite{wang2018cat},
for the same set of ISCAS-85 benchmarks,
we achieve 1.21$\times$ accuracy when splitting on M1
and 1.12$\times$ accuracy when splitting on M3 with less than 1\% running time.

\end{abstract}

\begingroup
\def\UrlFont{\ttfamily}
\maketitle
\endgroup

\section{Introduction}
Hardware becomes as vulnerable as software
with the widespread globalization of design, synthesis, fabrication,
and distribution of integrated circuits (ICs).
Fabless IC companies have to rely on external,
off-shore foundries for cost-effective access to advanced technology nodes,
which enables numerous attack avenues on intellectual property (IP)
including IP piracy~\cite{shamsi2019approximation},
tampering with hardware monitors~\cite{10.1007/978-3-662-49890-3_10},
and insertion of hardware Trojans~\cite{li2018practical}.

The IARPA agency advocated split manufacturing to safeguard chip designs
from potentially malicious foundries by splitting the designs
into the front-end-of-line (FEOL) and back-end-of-line (BEOL) parts~\cite{mccants11}.
An untrusted,
high-end foundry fabricates the device layer and a few lower metal layers,
whereas a trusted, low-end facility, which is possibly in-house,
integrates the BEOL on top of the FEOL by manufacturing the remaining metal layers~\cite{bi2015beyond}.
However,
merely splitting the designs into FEOL and BEOL may fall short in terms of security, as discussed next.

Traditional security-oblivious physical design tools place components
close to each other in the FEOL layers when they are connected
at the BEOL layers~\cite{li2018routability}.
While delivering effective designs in terms of power, performance, and area (PPA),
such an approach leads to some information leakage so that,
with an understanding of physical design tools,
the structural information gathered from the FEOL layers can be utilized to infer
the missing BEOL connections.
This generic concept is known as a \emph{proximity attack}.
Rajendran~\emph{et al.}~\cite{rajendran2013split} demonstrated the first na\"{i}ve proximity attack,
where they leveraged the fact that interconnected modules are typically placed close by.
The attack performed reasonably well for hierarchical designs with few nets
between the modules but showed limited success for flat designs and large layouts.
Wang~\emph{et al.}~\cite{wang2018cat} proposed
a network-flow model by setting the proximity as cost and capacitance as capacity.
However, the attack incurs significant runtime and the network flow is relaxed
to the na\"{i}ve proximity attack when cell libraries has loose capacitance constraints.
Zhang~\emph{et al.}~\cite{Zhang:2018:ASS:3195970.3195991}
analyze the security of split manufacturing on industrial designs with random-forest classifier.
However, their classifiers do not predict the BEOL connections directly,
but generate a list of candidates with considerable size instead.
For instance, when attacking layouts split on Metal 4 (M4), i.e., the FEOL contains M1--M4,
their most successful classifier provides on average several hundreds
or even thousands of candidates for each broken connection.
Here it can become practically impossible to retrieve all correct connections.

Besides traditional classifiers, we believe that deep learning (DL) is a good match
for attacking split manufacturing because attackers may have access
to an extensive database of designs to train on.
Among other applications, DL has been used to increase the effectiveness
and efficiency in human-level control~\cite{mnih2015human},
object recognition~\cite{he2017mask},
design-for-manufacturability (DFM)~\cite{yang2018layout,yang2018gan},
and routability prediction~\cite{Xie:2018:RRP:3240765.3240843}.
We caution that attacking split manufacturing with DL entails handling
a large variety of data.
Vector-based data are ranging, e.g., from signed gate displacements to unsigned wirelengths
and from integral pin counts to floating point pin capacitances.
Additionally,
layout images can represent routing segments and their directions as well as congestions.
Such image-based data naturally constitute rich information that can be useful
for an advanced attack.
One limitation of current neural network architectures is that
they can only handle either vector- or image-based features, but not both types together.
Accordingly, one challenge for our work is to combine both vector- and image-based features
in a unified framework.
Another limitation when applying DL for connection predicting is
that traditional two-class classifiers can only predict the probability of
each possible BEOL connection.
However, in the physical-design reality, each sink is assigned to exactly one source.
When selecting the source with the largest predicted connection probability,
traditional classifiers can be easily misled by outlying predictions of negative samples.
Furthermore,
traditional classifiers also suffer from the imbalance between positive and negative samples.

In this paper, we leverage deep learning to thoroughly learn the characteristics of
chip layouts arising from their physical design.
To the best of our knowledge,
this is the first deep-learning-based attack on split manufacturing
that provides better results
than the state-of-the-art non-learning-based attacks~\cite{wang2018cat}.
The major contributions of our work are as follows:
\begin{itemize}
    \item We leverage deep learning (DL) for attacking split manufacturing.
    We design and train a sophisticated deep neural network architecture
    which can predict the missing BEOL connections for unknown FEOL layout with high accuracy.
    \item Our neural network makes use of vector-based and image-based layout features
    simultaneously.
    \item The proposed \textit{softmax regression loss} allows our attack to directly
    and effectively select the best (i.e., most probable) BEOL connection
    among the relevant candidates.
\end{itemize}

The rest of the paper is organized as follows.
\Cref{sec:pre} outlines the threat model and problem formulation.
\Cref{sec:features} contains our features for the DL attack.
In \Cref{sec:dnn}, we describe the architecture and configuration of the proposed deep neural network
whose effectiveness is verified in~\Cref{sec:result}.
\Cref{sec:conclu} concludes the paper.

\section{Preliminaries}
\label{sec:pre}

\subsection{Threat Model}
Consistent with prior work~\cite{wang2018cat, Zhang:2018:ASS:3195970.3195991},
we assume that attackers have access to the full design information of the FEOL layers.
Hence, they can identify the gates, FEOL routing, and the resulting but incomplete netlist.
Also, they know the maximum load capacitances from the cell library
and can estimate an upper bound for the delay.
Further, consistent with prior work,
we assume attacks take place while chips are being fabricated.
Hence, oracle access is not available for the attacker.
Although we acknowledge a recent work~\cite{chen2018_SM_SAT}
where an oracle was leveraged to assist during the attack on split manufacturing,
here we adopt a stronger threat model where the chip is either
yet to be manufactured or is not available at all in the market.
We also assume that an attacker has a database of layouts generated in a similar manner
as the one under attack.

The objective of an attacker residing in the untrusted FEOL foundry is to decipher the missing BEOL interconnects solely from the available FEOL information.
The corresponding goal is to reconstruct the design and ultimately to pirate the chip IP,
overproduce the chip, or insert hardware Trojans.

\subsection{Problem Formulation}
\label{sec:terms}

\begin{figure}[tb]
\centering
\includegraphics[width=.9\columnwidth]{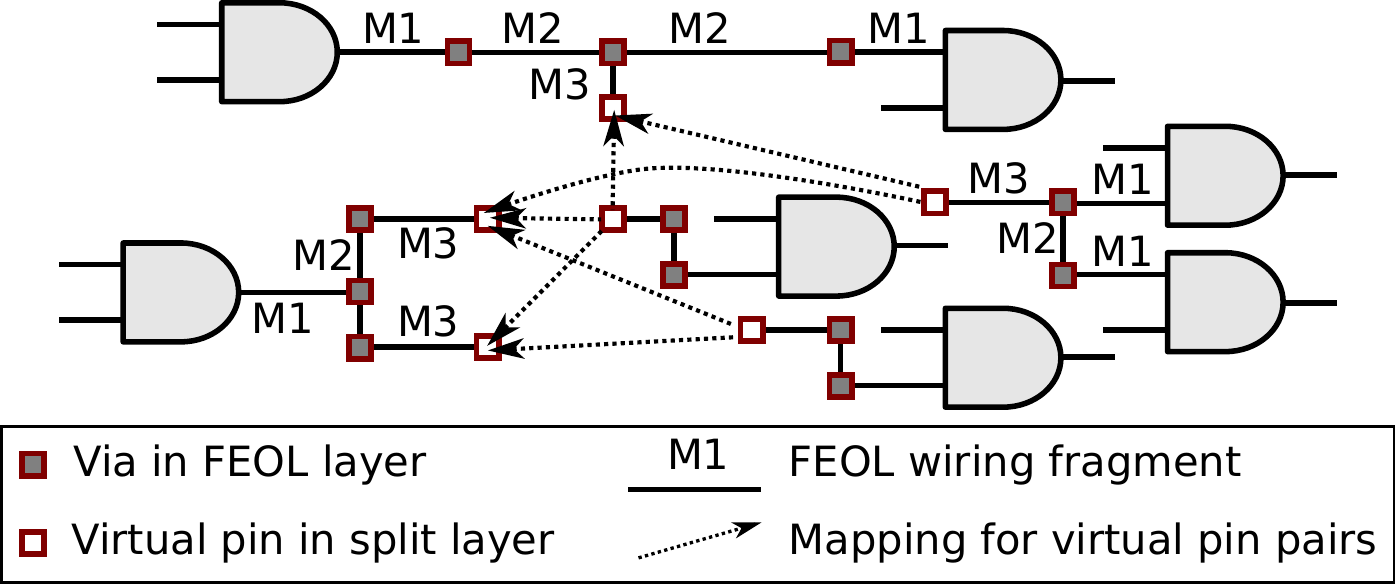}
\caption{
    Terms for learning on split manufacturing layouts.
    Examples of virtual pin pairs are shown by dashed arrows.
\label{fig:fragments}
}
\end{figure}

\emph{Split layer} is the top-most FEOL layer,
while \emph{virtual pins} are vias manufactured to connect the FEOL
with the BEOL~\cite{Zhang:2018:ASS:3195970.3195991}.
During split manufacturing, \emph{fragments} are connected parts of FEOL wires,
holding at least one virtual pin in the split layer.
There are two different types of fragments as shown in~\Cref{fig:fragments}: \begin{itemize}
    \item \emph{source fragment}: a driver/source along with outgoing fragments
    which are routed up until and within the split layer;
    \item \emph{sink fragment}: an incoming fragment which is routed within the split layer
    and down towards the sink(s).
\end{itemize}
For multi-fanout nets,
the sinks may be routed together in the FEOL as one sink fragment or
separately as several sink fragments.

Given a set of $m$ sink fragments, each of which has $c_1, c_2, \ldots, c_m$ sinks,
and a set of source fragments.
\emph{Virtual pin pairs} (VPPs) are mappings between sink fragment virtual pins
and source fragments virtual pins.
A VPP which is truly connected in the BEOL is called a \emph{positive VPP}.
Otherwise, it is called a \emph{negative VPP}.
The connection predicting problem is to select a VPP for each sink fragment
maximizing the correct connection rate (CCR)~\cite{wang2018cat},
which can be thought of as the percentage of sink pins that are successfully restored.
We define CCR as follows: \begin{equation}
    CCR = \frac{\sum_{i = 1}^{m} c_i x_i}{\sum_{i = 1}^m c_i},\label{eq:ccr}
\end{equation} where $x_i = 1~(0)$ when a positive (negative) VPP is selected
for the $i$-th sink fragment.

\section{Feature Extraction}
\label{sec:features}

\begin{figure*}[tb!]
    \begin{minipage}[]{.56\linewidth}
        \centering
        \subfloat[]{\includegraphics[width=.40\linewidth]{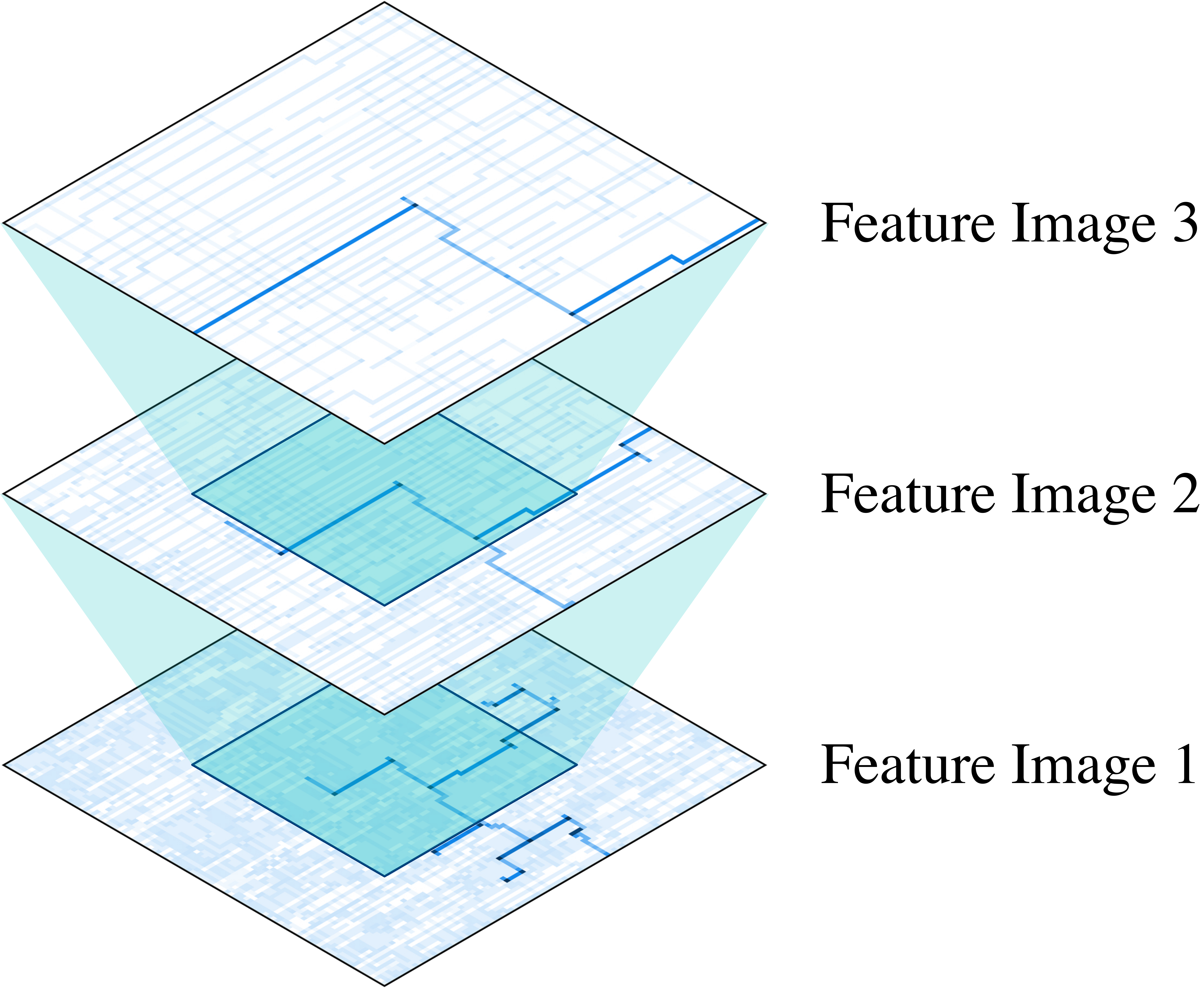} \label{fig:scale}}
        \subfloat[]{\includegraphics[width=.60\linewidth]{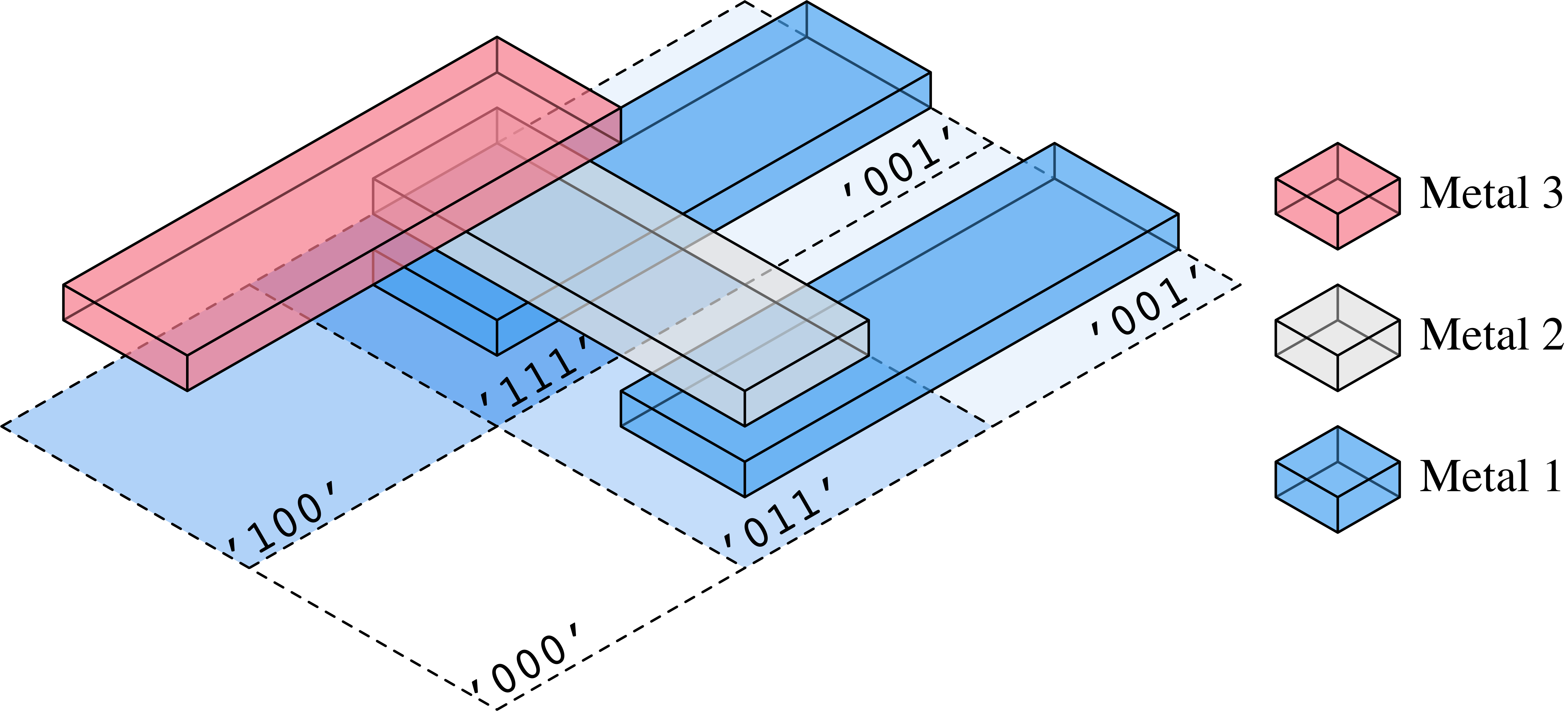}     \label{fig:image}}
        \caption{(a) Layout image scaling; (b) Layout image representation.}
    \end{minipage}
    \hspace{.1in}
    \begin{minipage}[]{.36\linewidth}
        \centering
        \includegraphics[width=1.0\linewidth]{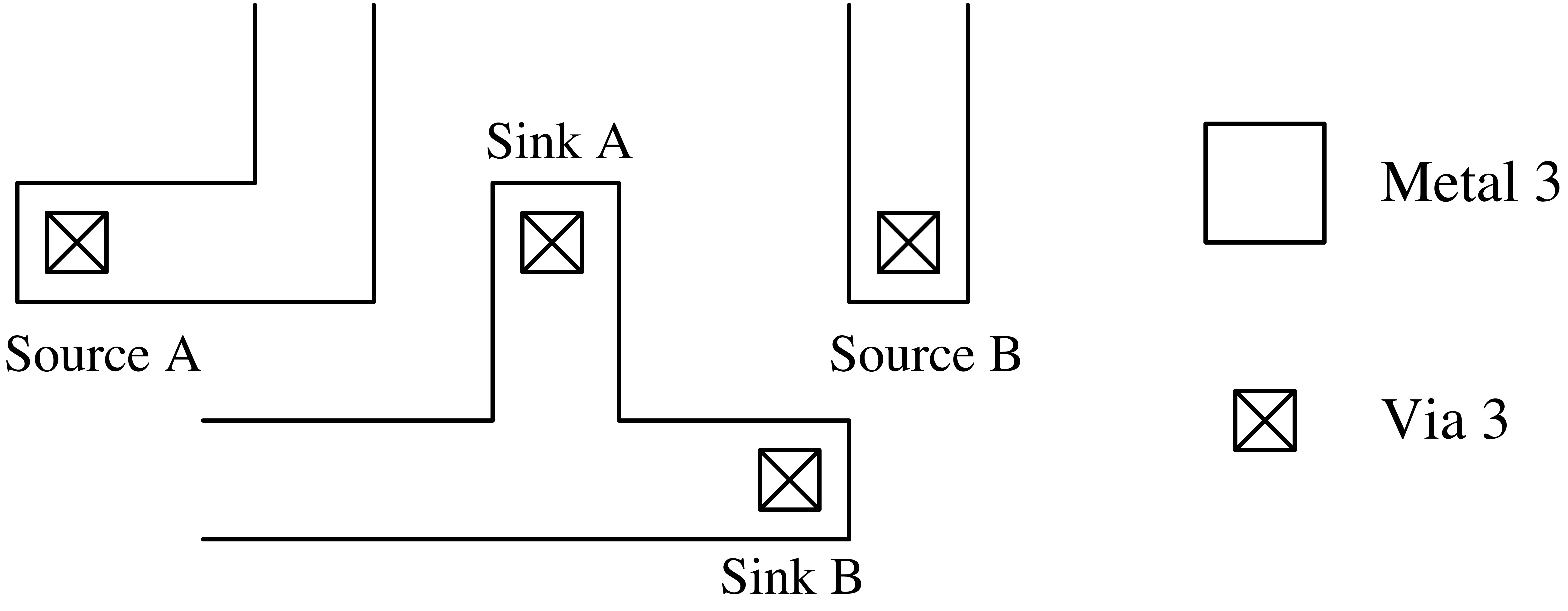}
        \caption{Direction example: Except VPP Source A--Sink B,
        all other VPPs are considered as candidates.
        \label{fig:dir}
    }
    \end{minipage}
\end{figure*}

The BEOL is only available at training time, where the true connectivity is extracted
to label VPPs as positive or negative ones.
The FEOL is available for both the training and attacking phases.
Hence, all features have to be tailored for the FEOL.
We propose two feature categories for our DL attack: vector- and image-based features.
We explain how to integrate these heterogeneous features
into a unified neural network architecture in the next section.

\subsection{Vector-based Features}

\subsubsection{Various Distances for VPPs}
These features capture the working essence of physical-design tools,
since gates to be connected are typically placed closer to each other.
In accordance to routing principles, all distances are captured as Manhattan distances.
For a VPP, the distances along the preferred and the non-preferred routing direction
as well as their sum are all considered separately.
Both signed and unsigned values are used to consider both the absolute and relative distances.
To mitigate scaling issues across layouts,
all distances are duplicated by encoding in the ratios of the chip width/height
and half-perimeter, respectively.

\subsubsection{Load Capacitance and Number of Sinks}
These features track the total load value and the number of sinks for each VPP.
As we are handling split or incomplete layouts,
the load capacitances can only be defined by two bounds as follows:
\begin{itemize}
\item upper bound: maximum capacitance of the driver, as derived from the cell library
(which is available to the attacker);
\item lower bound:
pin capacitance of the sinks connected within the sink fragment,
plus wire capacitances of the two related source and sink fragments.
\end{itemize}

\subsubsection{FEOL Layer Wirelengths and Vias}
These features capture the wirelength contribution in each FEOL metal layer individually.
Contributions are tracked separately for the two fragments of a VPP.
Within each layer, all wire paths of a fragment are summed up.
The number of vias in each FEOL cut layer is also considered.

\subsubsection{Driver Delay}
For each VPP,
we track the driver delay based on the underlying timing paths.
Note that timing paths obtained from split layouts can only provide lower bounds for delays,
as the paths may be incomplete.
Thus, this feature tends to become more meaningful for higher split layers,
when more of the paths are already completed in the FEOL.

\subsection{Image-based Features}
For each virtual pin,
we represent the routing layout of the local regions centering the virtual pin
as gray-scale layout images.
We consider three different scales with the same image shape
but different precisions as shown in~\Cref{fig:scale}.
Each image is 99 pixels wide and high, representing $99 \times 99$ consecutive regions.

There are two properties of the routed wires which will be encoded in the layout images
of a virtual pin:
the nets they belong to and the layers they are routed on.
Let $m$ be the number of metal layers in the FEOL.
The total number of bits in a pixel to represent the layout information is $2m$,
and we call these bits \textit{layer bits}.
$2m$ layer bits are needed because wires of the same fragment as the virtual pin
and wires from all other fragments are to be represented by different layer bits;
the more significant $m$ bits represent the routed wires of the virtual pin's fragment and
the less significant $m$ bits represent the wires of other fragments.
Wires of each type may exist in any routing layer of the FEOL,
and wires on different layers are also represented by different layer bits.
Since wires closer to the BEOL carry more information about the connection,
those in higher metal layers are encoded in more significant bits
while those in lower metal layers are encoded in less significant bits.
Vias connecting two layers are represented in both layer bits.

More specifically, a 1 is assigned to the $b$-th bit with $b = m, \ldots$, $2m - 1$ in a pixel
if the virtual pin's fragment is routed in metal layer $b - m + 1$ in that region.
Similarly, a 1 is assigned to the $b$-th bit with $b = 0, \ldots, m - 1$ in a pixel
if there is some wire or via arising from other fragments in metal layer $b + 1$ of that region.
\Cref{fig:image} shows part of the image data which is representative for a part of the layout
as in~\Cref{fig:fragments}, where the split layer is M3,
so that there may be wires on three different layers.
Routed wires in the six consecutive regions bounded by the dashed lines
are encoded into $2 \times 3$ pixels.
Note that here we only show the values of the sixth, fifth and fourth layer bits
at the corner of each region which together represent an exemplary virtual pin's fragment.

\section{Learning Framework}
\label{sec:dnn}
In this section, we start by describing the VPP selection for data cleaning.
We then demonstrate the neural network architecture and
discuss our proposed softmax regression loss and its advantages.

\subsection{Sample Selection}
\label{sec:dnn:sel}
Due to an underlying tendency towards imbalanced datasets
and long inference runtime,
it is not practical to consider all possible VPPs
because the correct connections are very few among all possible ones,
which leads to a biased or inaccurate ML model.
For $N$ nets,
even in the simplest scenario where ($i$) each FEOL wiring fragment holds only one virtual pin
in the split layer and ($ii$) each net is split into exactly one source and one sink fragment,
the sampling size is already $N^2$, whereas only $\frac{1}{N}$ samples are true positives.

\begin{table}
    \caption{VPP Preferences}
    \label{tab:prefer}
    \begin{tabular}{cc|ccc}
        \toprule
        Sk & Sc & Sk Prefers Sc & Sc Prefers Sk & Direction Criterion \\
        \midrule
        $A$ & $A$ & \ding{51} & \ding{55} & \ding{51} \\
        $A$ & $B$ & \ding{51} & \ding{51} & \ding{51} \\
        $B$ & $A$ & \ding{55} & \ding{55} & \ding{55} \\
        $B$ & $B$ & \ding{51} & \ding{51} & \ding{51} \\
        \bottomrule
    \end{tabular}
\end{table}

We select $n$ candidate VPPs for each sink in training and testing based on three criteria.
The first is the direction criterion.
We apply a looser criterion than~\cite{wang2018cat} to avoid neglecting positive VPPs,
based on our observation that wires with non-preferred routing direction
are relatively common in congested designs.
For a VPP $(p, q)$,
if $q$ is on the opposite side of one of the wire segments directly connected to $p$,
we then say the virtual pin $p$ \textit{prefers} virtual pin $q$.
The preferences of the VPPs in~\Cref{fig:dir} are shown in~\Cref{tab:prefer}.
Note that preferences of virtual pins in the same fragment are evaluated separately.
Now,
our direction criterion is that a VPP is \emph{not} considered as a candidate
in case both source and sink pins \emph{do not} prefer each other.
The second criterion is non-duplication.
If a sink fragment or source fragment have multiple virtual pins,
for each pair of sink fragment and source fragment,
only the VPP with the shortest distance apart in the non-preferred
routing direction of the split layer is considered as candidate.
This is based on the domain knowledge that the net wirelengths are restricted
to meet timing closure.
The third criterion is distance.
If the number of VPPs remaining is greater than $n$,
the VPPs with shorter distance in the non-preferred routing direction of
the split layer have higher priority to be selected.
If multiple VPPs tie on the non-preferred routing direction distance,
the distance in the preferred routing direction is considered as a tie-breaker.

\subsection{Model Architecture}
\label{sec:dnn:arch}
\begin{figure}[tb]
    \centering
    \includegraphics[width=\columnwidth,page=1]{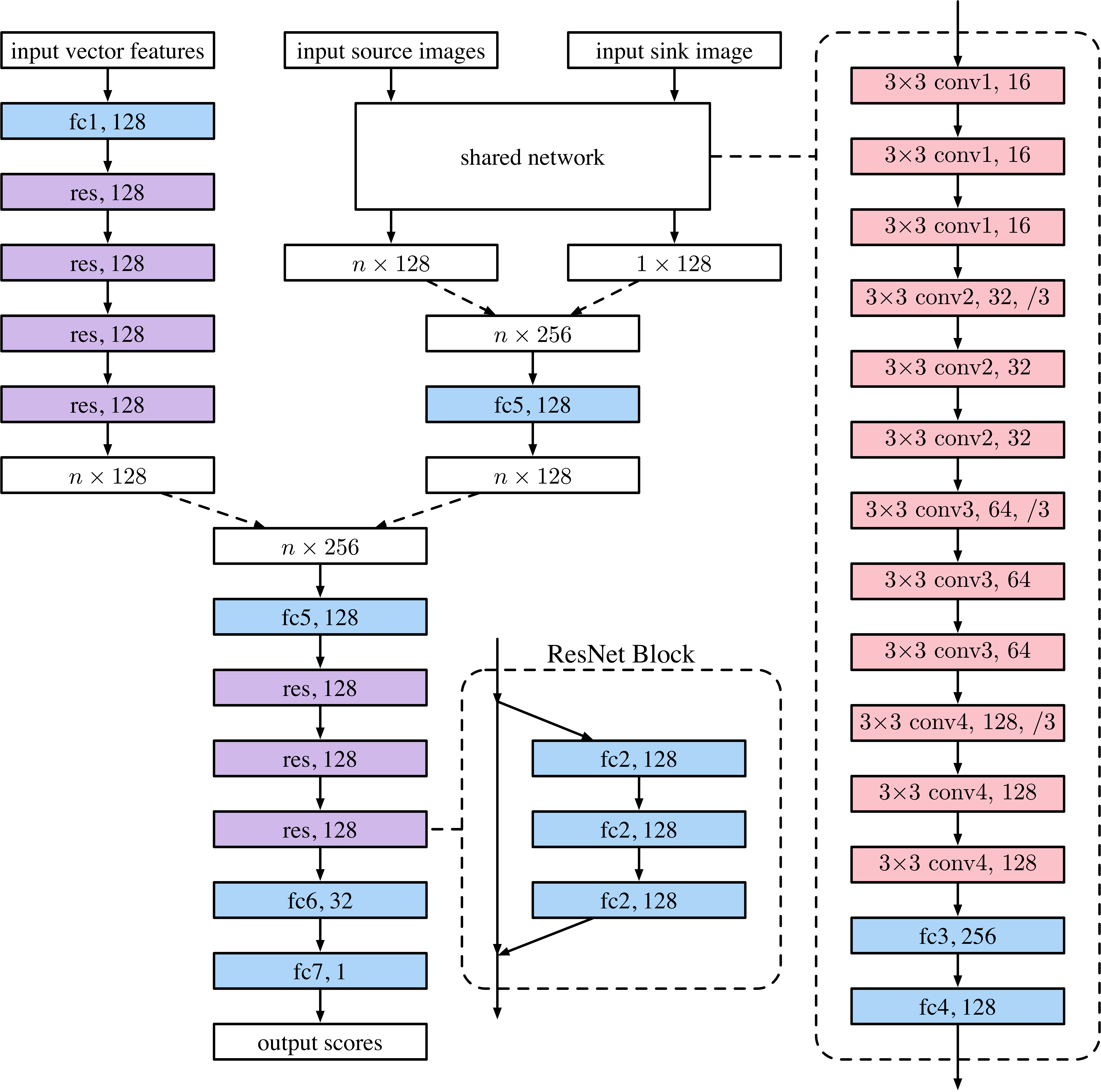}
    \caption{Neural Network Architecture.\label{fig:conv}}
\end{figure}

The input data for the model is a batch of features corresponding to a sink fragment
including the vector-based features of $n$ selected VPPs with the sink fragment,
the image-based features of $n$ source fragments in the related VPPs,
and the image-based features of the sink fragment itself.
The output data are scores for every VPPs in the batch.
To handle vector- and image-based features in the same network,
the proposed neural network illustrated in~\Cref{fig:conv}
first extracts underlying features from heterogeneous input by
processing vector-based features (shown in the upper left)
and image-based features (shown in the upper middle) individually,
and then processing them together (shown in the lower left)
after concatenating the output of the vector and image part together.

For the image part of the network,
note that the image-based features of the sink fragment are the same in the batch,
so we only process them once
to save runtime and its output is distributed to the output of every source images.
Besides, all the image-based features go through the same shared network.
The shared network contains twelve convolution layers (red colored, labeled as \texttt{conv})
and two fully connected layers (blue colored, labeled as \texttt{fc}).
The output from the sink image is then concatenated with every output from the source images
and the combination passes through one more 128-way fully connected layer.
For the vector part of the network,
vector-based features are first transformed by a 128-way fully connected layer.
Then, there are four residual networks (ResNet)~\cite{he2016deep} blocks
(purple colored, labeled as \texttt{res}).
The output of a ResNet block is the sum of its input and the output of
three fully connected layers as shown in the middle sub-figure of~\Cref{fig:conv}.
ResNet can resolve the gradient vanishing problem while training very deep neural networks.
After that,
the output from the image part is concatenated with the output from the vector-based features.
There is one 128-way fully connected layer to down-size the combination.
The network ends with three ResNet blocks and two more fully connected layers.
The filter and parameter configuration of the neural network is listed in~\Cref{tab:config}.
Both fully connected layers and convolutional layers are
followed by a leaky rectified linear unit (LReLU) $y = \max(0.01 x, x)$ as activation,
where $x$ is the input and $y$ is the output~\cite{maas2013rectifier}.

\begin{table}
    \caption{Neural Network Configuration.}
    \label{tab:config}
    \begin{tabular}{c|c|cc}
    \toprule
    Part & Layer & Parameter & Output \\
    \midrule
    Vector & fc1 & $27 \times 128$ & $n \times 128$ \\
    part & fc2 & $[128 \times 128] \times 12$ & $n \times 128$ \\ \midrule
    & conv1 & $[3 \times 3, 16] \times 3$ & $(n + 1) \times 99 \times 99 \times 16$ \\
    & conv2 & $[3 \times 3, 32] \times 3$ & $(n + 1) \times 33 \times 33 \times 32$ \\
    Image & conv3 & $[3 \times 3, 64] \times 3$ & $(n + 1) \times 11 \times 11 \times 64$ \\
    & conv4 & $[3 \times 3, 128] \times 3$ & $(n + 1) \times 4 \times 4 \times 128$ \\
    part & fc3 & $128 \times 256$ & $(n + 1) \times 256$ \\
    & fc4 & $256 \times 128$ & $(n + 1) \times 128$ \\
    & fc5 & $256 \times 128$ & $n \times 128$ \\ \midrule
    & fc5 & $256 \times 128$ & $n \times 128$ \\
    Merged & fc2 & $[128 \times 128] \times 9$ & $n \times 128$ \\
    part & fc6 & $128 \times 32$ & $n \times 32$ \\
    & fc7 & $32 \times 1$ & $n \times 1$ \\
    \bottomrule
    \end{tabular}
\end{table}

\subsection{Softmax Regression Loss}
Given a query of a batch of $n$ VPPs with at most one positive VPP,
the network predicts the connection probability $s_1, s_2, \ldots, s_n$ for each VPP.
The connection predicting task is to determine the index of the correct VPP to be connected:
\begin{equation}
    \arg \max_i s_i, \label{eq:out}
\end{equation}
as there can only be one source in a net.
Simply modeling the VPP connection problem as a two-class classification is not appropriate.
The main difference between our connection predicting problem
and the classical regression problems is that
we only care about the relative predicted probability between the only one positive VPP
and the remaining negative ones, instead of their absolute values.
Consequently, only the VPP with the largest predicted probability matters in the result.
An outlying negative VPP prediction would easily mislead the matching.

Considering a two-class classification formulation,
where the input of the neural network contains $n$ VPPs with the same sink fragment,
the loss of the two-class classification is \begin{equation}
    l_r = -\frac{1}{n}\left(\log\frac{e^{s_t^+}}{e^{s_t^-} + e^{s_t^+}} + \sum_{j \ne t} \log \frac{e^{s_j^-}}{e^{s_j^-} + e^{s_j^+}}\right),
\end{equation} whose partial derivative with respect to each score of either class is
\begin{equation}
    \frac{\partial l_r}{\partial s_j^+} = - \frac{\partial l_r}{\partial s_j^-} = \begin{cases}
    -\frac{e^{s_j^-}}{n\left(e^{s_j^-} + e^{s_j^+}\right)} & \text{if } j = t,\\
    \frac{e^{s_j^+}}{n\left(e^{s_j^-} + e^{s_j^+}\right)} & \text{otherwise,}
    \end{cases}
\end{equation}
where $s_j^+$ and $s_j^-$ are the scores of connection and non-connection
for the $j$-th source fragment with $1 \le j \le n$ and $t$ is the index of the true connection.
The partial derivative with respect to the $i$-th weight of either neuron
in the last fully connected layer is
\begin{equation}
    \frac{\partial l_r}{\partial w_i^+} = -\frac{\partial l_r}{\partial w_i^-} = \frac{1}{n}\left(\sum_{j = 1}^{n} \frac{e^{s_j^+} x_{i, j}}{e^{s_j^-} + e^{s_j^+}} - x_{i, t}\right),
\end{equation} where $x_{i, j}$ is the $i$-th input value of the last fully connected layer
for the $j$-th source fragment.
Therefore, the score of each source fragment acts independently on the gradient.
The coefficient of the positive part of the gradient, which is due to the negative samples,
is limited to 1 so that the VPP with even the largest connection probability
will not dominate the gradient.
As a result, misprediction of one VPP,
which significantly influences our desired output as in~\Cref{eq:out}, barely affects the average loss.
Additionally,
the numbers of positive and negative VPPs are imbalanced
as most of the VPPs are negative samples.
The negative part of the gradient, which is due to the only positive sample,
is divided by the number of VPPs in the batch.
Therefore,
a two-class classification model has a serious imbalance problem
as it can easily gain a high accuracy by simply classifying all VPPs as negative,
which is meaningless.

To resolve these problems,
we consider only one score $s_j$ for the $j$-th source fragment with $1 \le j \le n$.
We propose the following \textit{softmax regression loss}
\begin{equation}
    l_c = -\log\frac{e^{s_t}}{\sum_{j = 1}^{n} e^{s_j}},
\end{equation} whose partial derivative with respect to each score of connection is
\begin{equation}
    \frac{\partial l_c}{\partial s_j} = \begin{cases}
    \frac{e^{s_j}}{\sum_{j = 1}^{n} e^{s_j}} - 1 & \text{if } j = t,\\
    \frac{e^{s_j}}{\sum_{j = 1}^{n} e^{s_j}} & \text{otherwise.}
    \end{cases}
\end{equation}
The partial derivative of our proposed loss with respect to the $i$-th weight of
the only neuron in the last fully connected layer is \begin{equation}
    \frac{\partial l_c}{\partial w_i} = \frac{\sum_{j = 1}^{n} e^{s_j}x_{i, j}}{\sum_{j = 1}^{n} e^{s_j}} - x_{i, t},
\end{equation} in which the two problems of the two-class classification model are resolved.
Firstly, the source fragment with higher score contributes
much more significantly in the gradient with an exponential factor.
Let $j_{max}$ be the index of the largest $s_j$.
As the positive part of the loss is dominated by $x_{i, j_{max}}$,
we have $\frac{\partial l_c}{\partial w_i} \approx x_{i, j_{max}} - x_{i, t}$.
Secondly, the summation of the coefficients in the positive part equals to
that of the negative part, so there is no imbalance issue.
With these two advantages,
the proposed softmax regression loss reflects better our way of
computing the output as in~\Cref{eq:out}, which is also supported by the empirical results.

\section{Experimental Results}
\label{sec:result}

We derive a total of 9 training and 5 validation designs
from the ISCAS-85~\cite{hansen1999unveiling},
MCNC~\cite{MCNC}, and ITC-99 benchmarks~\cite{corno2000rt}.
We use the academic NanGate Open Cell Library~\cite{knudsen2008nangate}
for all our experiments.
Synopsys Design Compiler (DC) is used for synthesis.
Cadence Innovus 16.15 performs the placement and routing.
With regards to the to-be-attacked layouts,
we use the same benchmarks as mentioned in~\cite{wang2018cat} to ensure a fair comparison.
All training, validation and to-be-attacked layouts are free from DRC violations.
Once the layouts are generated,
we export a Design Exchange Format (DEF) file which is then split after the first and third metal layer
as two experimental sets to evaluate the performance with different split layer.
We implement the feature extraction with C++ and train the model with Python
and TensorFlow~\cite{abadi2016tensorflow}.
We select 31 VPPs for each sink fragment as the input of our DL attack
based on the proposed criteria in~\Cref{sec:dnn:sel}.
If the positive VPP is included in the collection,
there are 30 negative VPPs and one positive VPP selected.
If the positive VPP is not included, the predicted connection will definitely be wrong
according to the criteria.
We set each pixel in the image-based features representing a region of
$0.05 \times 0.05 \mathrm{\mu m}^2$,
$0.1 \times 0.1 \mathrm{\mu m}^2$, and $0.2 \times 0.2 \mathrm{\mu m}^2$, respectively.
The learning rate is set as 0.001 and decayed to 60\% for every 20 epochs.
We run all experiments on a 64-bit Linux machine with Intel Xeon 2.2GHz CPUs
and an Nvidia Titan V GPU.
We set the maximum runtime as 100,000 seconds (more than 24 hours) for all attacks
and report the CCR (\Cref{eq:ccr}) as the accuracy metric.

\begin{table*}[tb!]
    \centering
    \renewcommand{\arraystretch}{.9}
    \caption{Comparison with State-of-the-art Attack
    }
    \label{tab:layer}
    \begin{tabular}{c|cccccc|cccccc}
        \toprule
        \multirow{3}{*}{Design} &\multicolumn{6}{c|}{Metal 1} &\multicolumn{6}{c}{Metal 3}\\ \cmidrule{2-13}
        & \multirow{2}{*}{\#Sk} & \multirow{2}{*}{\#Sc} & \multicolumn{2}{c}{CCR (\%)} & \multicolumn{2}{c|}{Runtime (s)}
        & \multirow{2}{*}{\#Sk} & \multirow{2}{*}{\#Sc} & \multicolumn{2}{c}{CCR (\%)} & \multicolumn{2}{c}{Runtime (s)} \\
        &          & &\cite{wang2018cat}&Ours&\cite{wang2018cat}&Ours
        &&&\cite{wang2018cat}& Ours&\cite{wang2018cat}& Ours \\ \midrule
        \texttt{b11}	& 738	& 296	& 9.05	& 10.03	& 1719.46	& 11.06	& 213	& 57	& 66.67	& 66.67	& 0.94	& 4.20\\
        \texttt{b13}	& 430	& 215	& 10.42	& 17.91	& 130.82	& 7.53	& 88	& 52	& 42.05	& 70.45	& 0.44	& 3.55\\
        \texttt{b14}	& 6338	& 2864	& N/A	& 8.57	& $> 100000$	& 77.62	& 2117	& 583	& 30.33	& 30.42	& 2576.42	& 16.08\\
        \texttt{b15\_1}	& 10176	& 3847	& N/A	& 5.79	& $> 100000$	& 130.30	& 4910	& 1235	& 26.42	& 24.24	& 38292.53	& 33.50\\
        \texttt{b17\_1}	& 32385	& 12479	& N/A	& 4.08	& $> 100000$	& 599.47	& 16190	& 4590	& N/A	& 19.03	& $> 100000$	& 157.61\\
        \texttt{b18}	& 84292	& 33703	& N/A	& 4.59	& $> 100000$	& 2861.27	& 32719	& 9359	& N/A	& 23.74	& $> 100000$	& 453.66\\
        \texttt{b7}	& 520	& 235	& 8.43	& 10.19	& 326.13	& 8.55	& 115	& 51	& 55.65	& 84.35	& 0.67	& 3.62\\
        \texttt{c1355}	& 403	& 226	& 9.90	& 12.41	& 151.22	& 7.65	& 77	& 32	& 89.61	& 97.40	& 0.50	& 3.53\\
        \texttt{c1908}	& 432	& 213	& 8.49	& 11.11	& 260.50	& 7.45	& 54	& 27	& 94.44	& 87.04	& 0.47	& 3.34\\
        \texttt{c2670}	& 803	& 428	& 6.32	& 9.46	& 2251.82	& 11.70	& 206	& 120	& 54.85	& 58.74	& 1.48	& 4.64\\
        \texttt{c3540}	& 1354	& 512	& 6.41	& 8.49	& 39187.25	& 17.55	& 452	& 124	& 54.87	& 51.11	& 7.39	& 5.42\\
        \texttt{c432}	& 231	& 121	& 11.26	& 8.23	& 15.62	& 5.29	& 43	& 21	& 76.74	& 86.05	& 0.37	& 3.35\\
        \texttt{c5315}	& 1919	& 847	& 7.50	& 9.33	& 94281.90	& 23.59	& 590	& 248	& 52.20	& 62.03	& 26.11	& 6.81\\
        \texttt{c6288}	& 4124	& 2160	& N/A	& 14.52	& $> 100000$	& 49.64	& 551	& 78	& 63.16	& 61.52	& 7.13	& 4.22\\
        \texttt{c7552}	& 2008	& 1108	& 12.10	& 11.11	& 48656.51	& 22.82	& 296	& 175	& 50.34	& 72.30	& 7.64	& 3.72\\
        \texttt{c880}	& 460	& 234	& 11.09	& 13.91	& 568.99	& 6.31	& 77	& 37	& 71.43	& 76.62	& 0.74	& 2.34\\\midrule
        Average	& 	& 	& 9.18	& 11.11	& 13889.37	& 10.67	& 	& 	& 59.20	& 66.35	& 2923.06	& 7.02\\
        Ratio	& 	& 	& 1.00	& \textbf{1.21}	& 1.000	& \textbf{0.001}	& 	& 	& 1.00	& \textbf{1.12}	& 1.000	& \textbf{0.002}\\
        \bottomrule
    \end{tabular}
\end{table*}

In the first experiment, we illustrate the CCR for our proposed attacking method
and the state-of-the-art method~\cite{wang2018cat}.
\Cref{tab:layer} lists the results.
We evaluate the success of the network-flow attack ourselves
using the binary released in~\cite{NFA_attack}
whose runtime exceeds the limit on several large designs.
The CCR of our DL attack outperforms that of the state-of-the-art by 1.21$\times$
and 1.12$\times$ when splitting on M1 and M3, respectively.
Our inference time (including feature extraction) is significantly shorter.
Note that in the calculation of average,
designs on which~\cite{wang2018cat} times out are excluded for fairness.

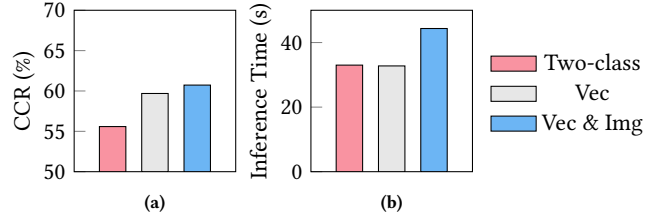
\begin{figure}[tb!]
    \centering
    \begin{tikzpicture}

\begin{groupplot}[
group style={
    group name=my plots,
    group size=2 by 1,
},
width=.44\linewidth,
height=.44\linewidth,
legend style={
    at={(0.5,1.1)},
    anchor=south,
    legend columns=-1,
},
    ybar=6pt,
    enlargelimits=0.05,
    enlarge y limits=0,
    legend style={at={(1.6,.8)},
    draw=none,anchor=north,legend columns=1},
    area legend,
    y label style={at={(0.26,0.5)}},
    symbolic x coords={\texttt{Average}},
    xtick=\empty,
    ytick pos=left,
]
\nextgroupplot[
bar width=10pt,
ylabel={CCR (\%)},
ymin=50,
ymax=70,
title style={at={(0.5,-0.4)}},
title={\footnotesize\textbf{(a)}},
]
\addplot[fill=OMNI_SALMON!80]    coordinates {(\texttt{Average},55.59)}; 
\addplot[fill=OMNI_MERCURY!80]     coordinates {(\texttt{Average},59.69)}; 
\addplot[fill=OMNI_RICH_SKY_BLUE!80]   coordinates {(\texttt{Average},60.73)}; 
\nextgroupplot[
bar width=10pt,
ylabel={Inference Time (s)},
ymin=0,
ymax=50,
title style={at={(0.5,-0.4)}},
title={\footnotesize\textbf{(b)}},
]
\addplot[fill=OMNI_SALMON!80]    coordinates {(\texttt{Average},33.01)};
\addplot[fill=OMNI_MERCURY!80]     coordinates {(\texttt{Average},32.76)};
\addplot[fill=OMNI_RICH_SKY_BLUE!80]   coordinates {(\texttt{Average},44.35)};
\legend{Two-class,Vec,Vec \& Img}
\end{groupplot}
\end{tikzpicture}
    \caption{
        Comparison between different settings of techniques used;
        (a) Compare the average CCR;
        (b) Compare the average inference time.
    }
    \label{fig:compare}
\end{figure}

In the second experiment,
we verify the effectiveness of our proposed softmax regression loss
and the image-based features.
As shown in~\Cref{fig:compare},
the baseline is splitting on M3 and using only the vector-based features with the loss function of
the two-class classification.
With the softmax regression loss, the average CCR is 1.07$\times$ that of the baseline.
With additional image-based features, the average CCR further improves to 1.09$\times$.
The use of softmax regression loss improves the runtime marginally.
The runtime when additionally using the image-based features is comparable to that of only
using the vector-based features.

\section{Conclusion}
\label{sec:conclu}
We presented an attack on split manufacturing using deep learning.
Firstly, we demonstrated vector-based and image-based features
and a neural network that can process these heterogeneous features simultaneously.
We further proposed a softmax regression loss
that directly reflects on the accuracy for the virtual pin pair matching problem of split manufacturing
and, thereby, eliminates imbalance issues found in prior art.
Compared with the state-of-the-art network-flow attack~\cite{wang2018cat},
the correct connection rate is improved by 1.21$\times$ and 1.12$\times$
when splitting on Metal 1 and Metal 3, respectively.
Moreover, the runtime for our attack is significantly less,
namely less than 1\% of that of~\cite{wang2018cat}.
As a part of future work, we plan to extend our neural network to tackle industrial layouts which have been incorporated with various
placement-based and/or routing-based defense strategies.

\begin{acks}
    This work is supported in part by \grantsponsor{}{The Research Grants Council of the Hong Kong Special Administrative Region, China}{} (Project No.~\grantnum{}{CUHK14202218}) and \grantsponsor{}{Center for Cyber Security Abu Dhabi (CCS-AD) in New York University Abu Dhabi}{}.
\end{acks}

\bibliographystyle{IEEEtran}
\bibliography{sample-bibliography}

\end{document}